\documentclass{jetpl}

\twocolumn
\def\refitem#1{\relax}

\usepackage{graphics} %

\lat

\newcommand{\nee}{n_e}
\newcommand{\nc}{n_c}
\newcommand{\RM}{{\rm RM}}
\newcommand\vect[1]{{\mathbf {#1}}}

\newcommand{\aap}{Astron. Astrophys.}
\newcommand{\mnras}{Mon. Not. R. Astron. Soc.}
\newcommand\apj{ApJ}
\newcommand\pre{Phys.\ Rev.\ E}
\newcommand\prd{Phys.\ Rev.\ D}
\newcommand\nat{Nat}

\title{Helicity detection of the astrophysical magnetic fields from radio emission statistics}
\rtitle{Magnetic helicity detection}

\sodtitle{Helicity detection of the astrophysical magnetic field from radio emission statistics}

\author{A. Volegova and R. Stepanov \thanks{e-mail: rodion@icmm.ru}}
\rauthor{A. Volegova and R. Stepanov}
\sodauthor{A. Volegova and R. Stepanov}

\address{Institute of Continuous Media Mechanics, Ac. Koroleva str. 1, 614013 Perm, Russia}

\dates{Received 29 September, 2009}

\abstract{We discuss inverse problem of detection turbulence magnetic field helical properties using radio survey observations statistics.
In this paper, we present principal solution which connects magnetic helicity and correlation between Faraday rotation measure and polarization degree of radio synchrotron emission. The effect of depolarization plays the main role in this problem and allows to detect magnetic helicity for certain frequency range of observable radio emission.
We show that the proposed method is mainly sensitive to a large-scale magnetic field component.
}


\begin{document}
\maketitle

\section{Introduction}
Magnetic fields exist not only in compact astrophysical objects such as planets and stars but also are observed everywhere in the Universe, interstellar space and can be attributed to galaxies and galactic clusters \cite{1997Natur.385..131Z}. Substantially the dynamo theory explains the nature and evolution of cosmic magnetic fields \cite{1979cmft.book.....P}.
The dynamics specific property of magnetohydrodynamics (MHD) systems are turbulent motions of media. The mean field theory \cite{1980opp..bookR....K} predicts magnetic field generation, as a result of $ \alpha $-effect which can appear under the condition of helical turbulent flows.
As a result magnetic field becomes helical too \cite{2001ApJ...550..824Brandenb} and can be described by magnetic helicity
 $H=\vect{A} \cdot \vect{B}$,
 where $ \vect {A} $ - is the vector potential of magnetic field $ \vect {B} = {\rm curl} \, \vect {A} $.

Recently the special role of magnetic helicity in space magnetic fields evolution processes is noted \cite{2007PPCF...49..447S}. Total magnetic helicity of a system is integral of motion and conserved in the nondissipative limit.
The results of theoretical and numerical researches show that the magnetic helicity can accumulate in the system and suppress the generation mechanisms \cite{2007PhRvE..76b6316M}. This put into question a possibility of the  turbulent dynamo and has demanded construction of the adequate model describing dynamics $H$.

The model of dynamo in the galactic disk has been added by equations describing outflux magnetic helicity that has allowed to overcome catastrophic suppression dynamo processes \cite{2006A&A...448L..33S}. Thus, mechanisms of solar dynamo have been considered and it is shown that allowance for the helicity of the small-scale magnetic fields is of crucial importance in limiting the energy of the generated large-scale magnetic field  \cite{Pipin2007AJ}.
Using the results of \cite{2006A&A...448L..33S}, it is proved necessity of coronal ejections for the strong large-scale solar magnetic field generation \cite{2007HiA....14..291B}.

The development of the models that describe evolution of magnetic helicity requires understanding of nonlinear processes in multi-scale systems as well as notions about magnetic energy and helicity spectral distributions, non-uniformity and anisotropy properties of the spatial distributions.
It is extremely important to have factual material confirming presence of helicity and its connections with other components of the media.

The observations of helicity in the solar convective zone indicate the existence of connection between the intensity of current helicity and dynamo processes \cite{2006MNRAS.365..276Z}.
Study of MHD turbulence in laboratory conditions is extremely difficult (for
review, see \cite{Stefani2008}).
Single successful experiments for measuring the turbulent magnetic fields \cite{Denisov08:JETPL} considerably differ in values of the characteristic parameters, primarily the magnetic Reynolds number.
Analysis of astrophysical observations remains the most promising direction of research in this question.

In current astrophysical researches there is no general approach to derivation helicity of interstellar magnetic fields.
It is discussed a possible way to detect magnetic helicity from cosmic microwave background (CMB) fluctuations data  \cite{2006NewAR..50.1015K}. In some cases \cite{2006PhRvD..73f3507K}, information about helicity can be extracted from the properties of cosmic rays if their source is known.
The authors of these researches noted that their approaches requires presence of high-accuracy observation data, which we don't have at present, therefore the practical application of these approaches is limited.

New generation of radio telescopes (SKA and LOFAR) offers great opportunities \cite{2007AdRS....5..399B} because new high accuracy and resolution data about the space magnetism will be available in the near future.
Magnetic fields of the interstellar medium are the most suitable object for derivation MHD turbulence properties.
Due to relatively large scales, we can neglect the contribution of regular magnetic fields, stars and planets and assume that the continuous electrically conductive interstellar medium is in a state of turbulent motion, which is excited by explosions of supernovae \cite{sokoloff88}.
Another significant feature of the interstellar medium is in its "transparency"{}, i.e. depth distribution of radio sources allows to make an analysis in all three dimensions.

The aim of the paper is to show the possibility of magnetic helicity detection in the ionized interstellar plasma by statistical analysis of radio polarized observations.
We consider the model distribution of the magnetic field with given properties and determine relation of  magnetic helicity and correlation coefficient between Faraday rotation measure and polarization degree of radio emission.

\section{Interstellar medium model}

The simulation domain is a cube of side $L$. Let the coordinates $x, y $ describe the sky plane and the axis $z $ corresponds to the line of sight.  For generation of artificial polarized radio data distributions it is necessary to define some distributions of the ISM components such as magnetic field $\vect{B}$, densities of relativistic $ \nc $ and free thermal electrons $ \nee $.
An indicator of the magnetic field in the interstellar medium is synchrotron emission resulting from relativistic electrons passing through the magnetic field.

At the first step of work we calculate three-dimensional distribution of uniform and isotropic magnetic field.
The input parameters of the model are power low of spectrum for magnetic energy distribution $ \alpha $, the turbulent energy scale $l $ and magnetic helicity value. Also, the resulting magnetic field is solenoidal, i.e. $ \vect {\nabla} \cdot\vect {B} =0$. We can satisfy this conditions conveniently using the Fourier representation of the magnetic field  $ \vect {\hat {B}} $ expressed via the vector potential $ \vect {A} $

\begin{equation}\label{B}
\vect{\hat{B}}(\vect{k})= i \vect{k}\times\vect{\hat{A}}(\vect{k}), \;\;\;
\vect{\hat{A}}(\vect{k})=
%\frac{\vect{c}}{|\vect{k}\times\vect{c}|}(|\vect{k}|/k_0)^{\alpha(k)/2-1}
\frac{\vect{c}}{|\vect{k}\times\vect{c}|}|\vect{k}|^{\alpha/2-1},
\end{equation}

where $ \vect {k} $ is the wave-vector, $ \vect {c} = \vect {a} +i \, \vect {b} $ is the random complex vector, whose distribution determines magnetic helicity value. If the random vectors $ \vect {a} $ and $ \vect {b} $ have uniform distribution the mean value of helicity $ \langle H\rangle $ will be close to zero. If we choose only those pairs of vectors which give the same sign of $ \vect {k} \cdot (\vect {a} \times\vect {b})$ then $\langle H\rangle $ will be positive or negative, respectively. The extreme value of magnetic helicity for given magnetic energy will be obtained with
\begin{equation}\label{c_H}
  \vect{b}= \pm \frac{\vect{k}\times\vect{a}}{|\vect{k}\times\vect{a}|}|\vect{a}|.
\end{equation}
The turbulent cells number along line of sight is defined by $N = [L k_{0}] $, where $k_{0} =l^{-1} $ is the length of the wave-vector till which magnetic energy is equal to zero. Starting from $k_{0}, $ the energy spectrum obeys the Kolmogorov law $ \alpha =-5/3$.

\begin{figure*}
\centering
\includegraphics[width=0.79\textwidth]{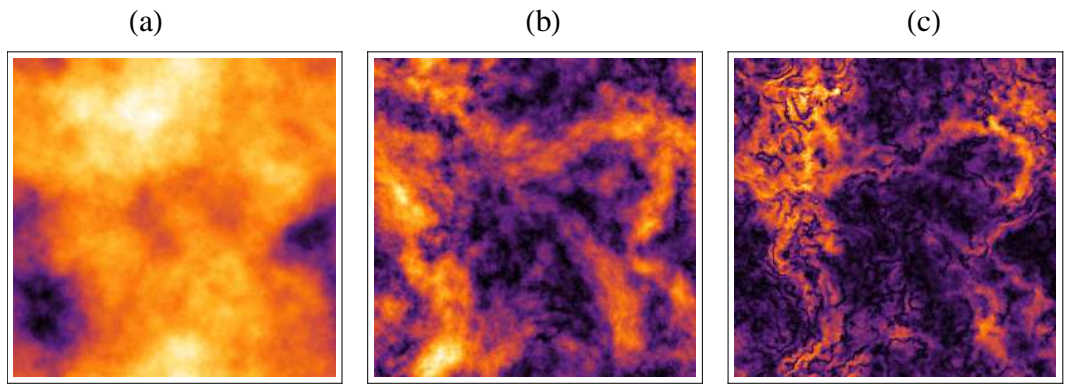}
\caption{
 Fig.1. Radio maps. Distributions are represented in the observation plane $(x,y)$: (a) Faraday rotation measure distribution $\RM(x,y,z=L)$, (b) Polarization degree $p$ for $\lambda = 0.05$\,m and (c) for $\lambda = 0.2$\,m. Image resolution is 256x256 pixels. Black colour corresponds to minimum values and white corresponds to maximum values}
\label{fig1}
\end{figure*}

The next step of solving the problem is calculation of artificial polarized radio emission maps and $\RM$ images. The total intensity of synchrotron emission  is given by
\begin{equation}\label{I}
    I(x,y)= \int_0^L \epsilon(x,y,z) dz,
\end{equation}
where $ \epsilon (x, y, z) $ is the synchrotron emissivity. Using simplified representations
about spectral distribution of $ \nc $ we can consider that
$ \epsilon\sim\nc (B_x^2+B_y^2) $. The synchrotron emissivity initially has some degree of polarization $ \gamma $ and the polarization
 angle is defined by a perpendicular direction to $ \vect{B} $ in the sky plane $ (x, y) $.
The intrinsic polarization angle at the point of emission $ (x, y, z) $ is given by
\begin{equation}\label{psi0}
    \psi_0(x,y,z)=\arctan(B_y/B_x)+\pi/2.
\end{equation}
When polarized radio emission propagates thought magnetized plasma, the polarization plane rotated by Faraday effect. Thus, the polarization angle at some point with \begin{equation}\label{psi}
    \psi(x,y,z)=\psi_0(x,y,z)+\lambda^2 \RM(x,y,z),
\end{equation}
where $ \lambda $ is the wavelength of observed radio emission, and $ \RM $ is Faraday rotation measure which determined by
the integral with variable upper limit
\begin{equation}\label{RM}
    \RM(x,y,z)=K \int_0^z \nee B_z(x,y,z') dz'.
\end{equation}
Note that Faraday rotation of the polarization plane depends on the magnetic field component along the line of sight,
 while polarized intensity and polarization angle are defined by the perpendicular component.
The observed Stocks parameters $Q $ and $U $ can be used to determine the complex intensity
of the polarized emission $P=Q+{\rm i} \, U $ which given as
\begin{equation}\label{P}
    P(x,y)= \gamma \int_0^L \epsilon(x,y,z) \exp{\{2\psi(x,y,z)\}} dz.
\end{equation}
Superposition of the electromagnetic waves with different polarization angles causes depolarization, thus that the observed polarization degree $p = |P |/I $ varies from 0 to $ \gamma $. Depolarization may be caused not only by physical reasons and also by the limited radio telescope resolution. In this work instrumental effects are not considered. The physical size of the simulation domain is $L=0.5$~kpc (1 kiloparsec $ \approx 3\cdot10 ^ {19} $ \,) that corresponds to the half-thickness of the galactic disk. The dimensional constant in (\ref{RM}) $K=0.81$ if $ \lambda $ is measured in meters, $z $ in parsecs, and $ \nee $ in ${\rm cm}^{-3}$.
We accepted the mean value of the magnetic field $\overline{B}=1$~$ \mu $G and the thermal electron density $\nee=1 {\rm cm}^{-3}$ as a typical values for the ISM.

Figure~1 shows typical view of calculated Faraday rotation measures and polarization degree distributions for wavelengths of radio emission 0.05~m, 0.2~m.
These distributions of the radio data contain information about all magnetic field components.
The changes of distribution details $p $ for long $ \lambda $  are explained by Faraday depolarization.
It is possible to see formation of thin black structures (figure~1c) in domains corresponding to the maximal values of Faraday rotation measure, this structures are typical for real astrophysical observations.
The analysis of these structures called "canals"{}, allows to identify some properties of interstellar turbulence \cite{2006MNRAS.371L..21F}. Noted canals appearance reflects the fact of reliability for the chosen ISM model.

\section{Statistical analysis}

Figure~2 shows probability distribution functions of $ \RM $ for three levels of magnetic helicity. These distributions have symmetric shapes which are sufficiently approximated by the normal law. The difference between distributions is insignificant that cannot be used for diagnostics of helicity.
\begin{figure}
\includegraphics[width=0.40\textwidth]{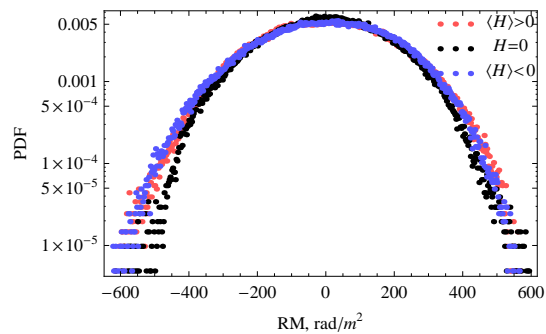}
\caption{Fig.2. Probability distribution function of Faraday rotation measure for different levels of magnetic helicity}
\label{fig2}
\end{figure}
\begin{figure*}
\label{fig3}
\centering{
\includegraphics[width=0.99\textwidth]{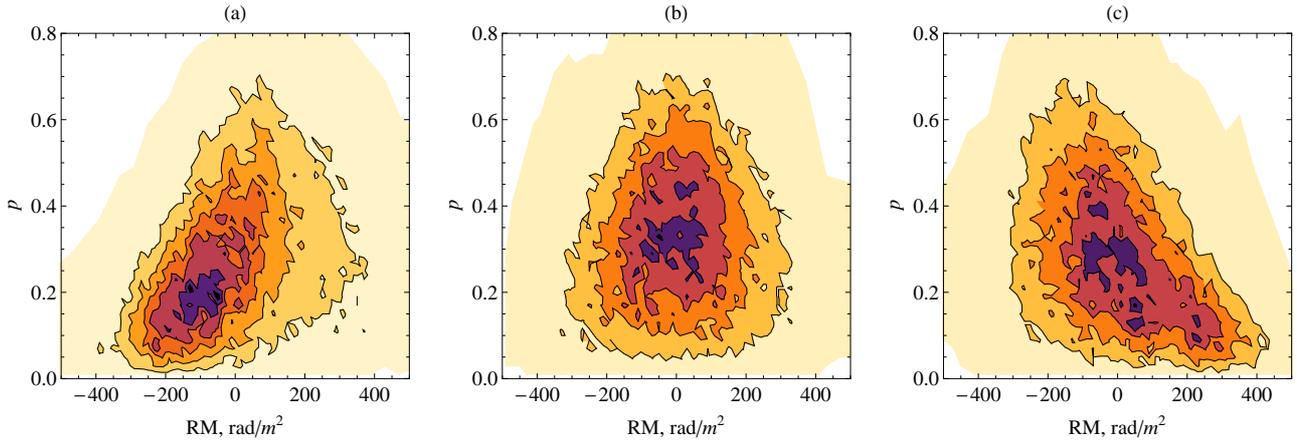}}
\caption{Fig.3. Joint probability distribution of pair $\RM$ and $p$ for three levels of magnetic helicity : (a) positive, (b) zero-order, (c) negative}
\end{figure*}

The situation changes substantially if we consider joint probabilities.
The joint probability distribution density of $ \RM $ and $p $ depending on sign of magnetic helicity is shown in Fig.~3.
Magnetic helicity destroys the symmetry of the distribution function.
For $ \langle H\rangle> 0$ (Fig.~3a) low degree of polarization is most likely corresponds to negative values of $ \RM $, and for $ \langle H\rangle <0$ (Fig.~3c) -- positive values.
The quantitative estimation of revealed statistical characteristics can be derived using the correlation coefficient
\begin{equation}\label{r}
  C= \frac{\langle RM p\rangle-\langle RM\rangle \langle p\rangle}
  {\sqrt{(\langle RM^2 \rangle-\langle RM\rangle^2)(\langle p^2 \rangle-\langle p\rangle^2)}},
\end{equation}
where mean value is taken in the observation plane $ (x, y) $ for $z=L $.
The probability distribution function of $ C $ is defined by repeated calculations of random magnetic field realizations with given level of magnetic helicity. For construction of the results that shown in Fig.~4 we used 300 realizations for each level of magnetic helicity, respectively.
The ranges of $ C $ values for each magnetic helicity level practically  doesn't intersect that allows to identify the sign of magnetic helicity.
As mentioned previously, the main reason for connection of $ \ RM $ and $ p $ is probably the depolarization effect caused by Faraday rotation, that confirms the relationship of mean correlation $\overline{C}$ with the wavelength
$\lambda $ (see Fig.~5).

\begin{figure}
\label{fig4}
\centering{
\includegraphics[width=0.40\textwidth]{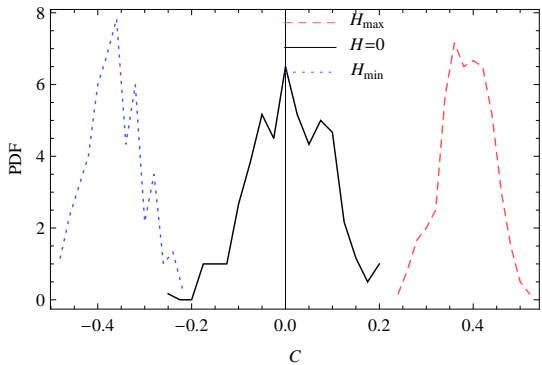}
}
\caption{Fig.4. Probability distribution function of $C$ with $\lambda=0.2$~m for three levels of magnetic helicity }
\end{figure}

\begin{figure}
\label{fig5}
\centering{
\includegraphics[width=0.40\textwidth]{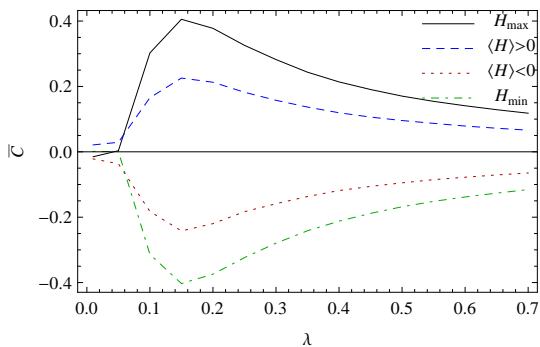}
}
\caption{Fig.5. Mean of correlation $\overline{C}$ depending on wave length $\lambda$ for different levels of magnetic helicity}
\end{figure}

For wavelengths less than 6~cm we have low Faraday depolarization that $ \overline{C}$ is almost equal to zero.
The most successful wavelength for given parameters of the interstellar medium model is $\lambda\approx15$~cm, which produces the extremum of $\overline{C} $.
The connection between $\overline{C}$ and the number of turbulent cells along the line of sight $N$ also was investigated. Our calculations show that the value of $\overline{C}$
suddenly decreases with increasing number of turbulent cells (see~Fig.6).

According to the mathematical modeling results, the magnetic helicity initiates a correlation between polarization degree and Faraday rotation measure. Thus strong correlation is approximately equal to 0.4 and achieved for a definite wavelength. Optimum wavelength for the observations will depend on the parameters of the interstellar medium: the domain size $L$, magnetic field $B$, the thermal electron density $\nee$. However, it is found that for maximum effect, the interstellar medium should provide specific Faraday rotation, rotation of the polarization plane through angle of $2\pi$. And then relation
\begin{equation}
\RM \,\lambda^2\approx K \,L \,B\, \nee\, \lambda^2\approx 2\pi.
\end{equation}
The correlation decreases with increasing of the number of turbulent cells along the line of sight. It means that the proposed method for diagnosis of the magnetic helicity is mainly sensitive to large-scale magnetic field component.
The noise appears during radio polarized observations process, probably, also produces accuracy problems for helicity detection. This influence can be assessed within the proposed model, but it makes sense to do so, when we have real data with known signal-to-noise ratio and other features.

\begin{figure}
\label{fig6}
\centering{
\includegraphics[width=0.40\textwidth]{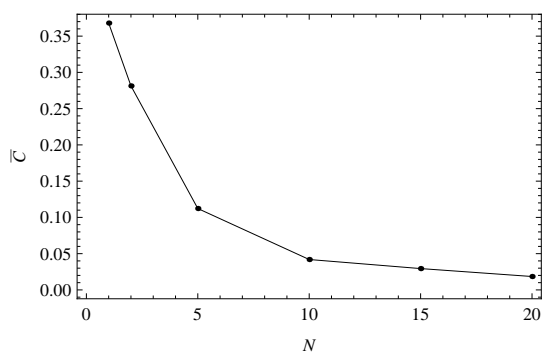}
}
\caption{Fig.6. Mean of $\overline{C}$ as a function of  turbulent cells number $N$ along the line-of-sight for $\lambda=0.2$~m}
% depending on
\end{figure}

This work was supported by the Russian Foundation for Basic Research (projects \#08-02-92881-DFG and \#07-01-96007-Ural) and the
International Science and Technology Centre (project \#3726).
The authors are grateful to A. Shukurov, D. Sokoloff and P. Frick for useful discussions.

%\bibliographystyle{maik} % this style correspond to lournal style
%\bibliography{ism}

\end{document}